\documentclass[aps,prl,twocolumn,superscriptaddress]{revtex4-1}
\newcommand{\revision}[1]{{{#1}}}

\usepackage{graphicx}% Include figure files
\usepackage{dcolumn}% Align table columns on decimal point
\usepackage{bm}% bold math
\usepackage{accents}
\usepackage{amssymb}
\usepackage{feynmp}
\usepackage{url}
\usepackage{setspace}
\usepackage{color}
\usepackage{hyperref}

\newcommand{\beq}{\begin{equation}}
\newcommand{\eeq}{\end{equation}}
\newcommand{\beqa}{\begin{eqnarray}}
\newcommand{\eeqa}{\end{eqnarray}}

\newcommand{\del}[2]{\frac{\dd #1}{\dd #2}}

\newcommand{\av}[1]{\left\langle #1 \right\rangle}

\newcommand{\dd}{\mathrm{d}}

\begin{document}

\title{Shortcuts to Adiabatic Pumping in Classical Stochastic Systems}

\author{Ken Funo}
\affiliation{Theoretical Physics Laboratory, RIKEN Cluster for Pioneering Research, Wako-shi, Saitama 351-0198, Japan}
\author{Neill Lambert}
\affiliation{Theoretical Physics Laboratory, RIKEN Cluster for Pioneering Research, Wako-shi, Saitama 351-0198, Japan}
\author{Franco Nori}
\affiliation{Theoretical Physics Laboratory, RIKEN Cluster for Pioneering Research, Wako-shi, Saitama 351-0198, Japan}
\affiliation{Physics Department, The University of Michigan, Ann Arbor, Michigan 48109-1040, USA}
\author{Christian Flindt}
\affiliation{Department of Applied Physics, Aalto University, 00076 Aalto, Finland}

\date{\today}

\begin{abstract}
\revision{Adiabatic pumping is characterized by a geometric contribution to the pumped charge, which can be non-zero even in the absence of a bias. However, as the driving speed is increased, non-adiabatic excitations gradually reduce the pumped charge, thereby limiting the maximal applicable driving frequencies. To circumvent this problem, we here extend the concept of shortcuts to adiabaticity to construct a control protocol which enables geometric pumping well beyond the adiabatic regime. Our protocol allows for an increase, by more than an order of magnitude, in the driving frequencies, and the method is also robust against moderate fluctuations of the control field. We provide a geometric interpretation of the control protocol and analyze the thermodynamic cost of implementing it. Our findings can be realized using current technology and potentially enable fast pumping of charge or heat in quantum dots, as well as in other stochastic systems from physics, chemistry, and biology.}
\end{abstract}

\maketitle

\begin{figure}[t]
\begin{center}
\includegraphics[width=.48\textwidth]{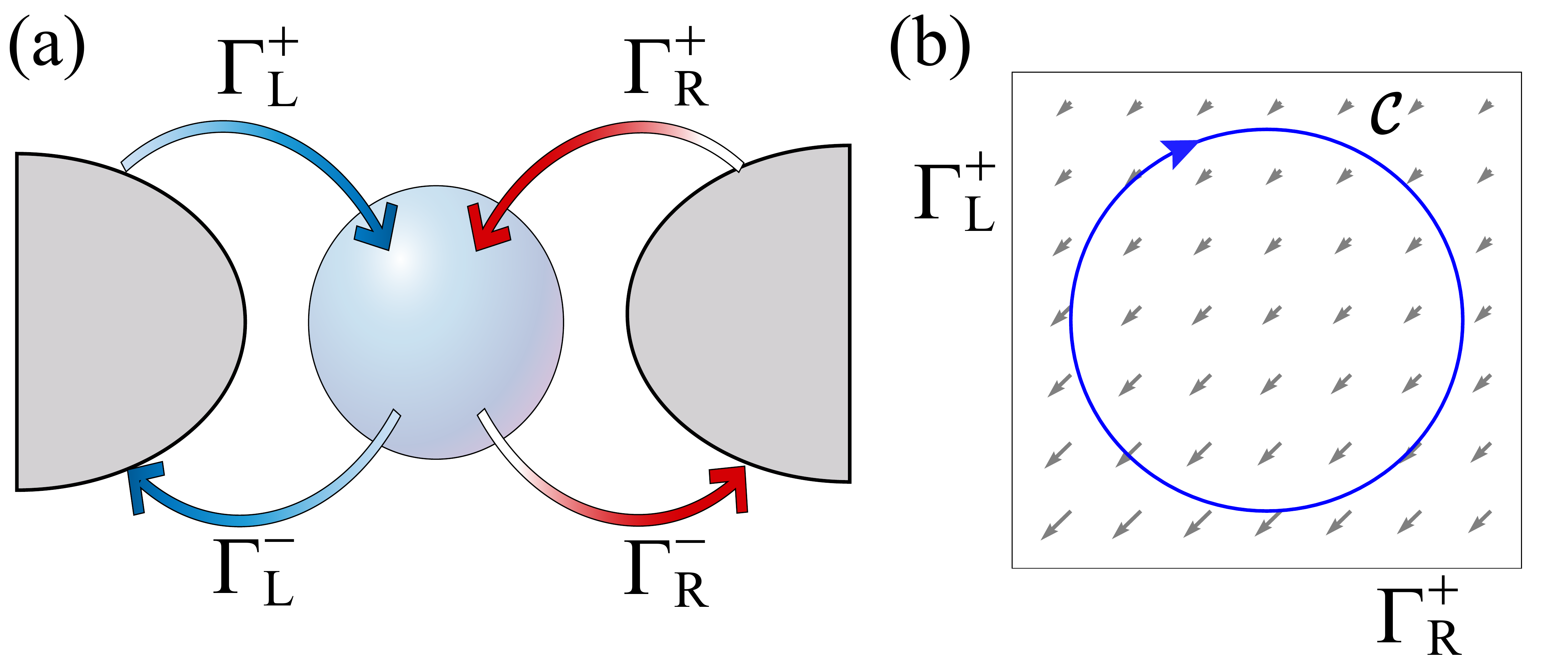}
\caption{\revision{Adiabatic pumping. (a) Adiabatic pumping can be implemented in a quantum dot with time-dependent tunneling rates. (b) At low driving frequencies and without a bias, the pumped charge can be expressed as an integral over the vector potential ${\bm A}({\bm \Gamma})$ along a closed contour $\mathcal{C}$ defined by the time-dependent rates ${\bm \Gamma}=\{\Gamma^{+}_{\rm R},\Gamma^{+}_{\rm L}\}$. The pumped charge is purely geometric, since it only depends on the contour $\mathcal{C}$.}}
\label{fig_rates}
\end{center}
\end{figure}

{\it Introduction.---} \revision{Adiabatic driving makes it possible to pump charge or heat by slowly modulating two or more system parameters periodically in time. Even without an applied bias, the slow driving can induce a non-vanishing pumped charge, which resembles the  Berry phase in quantum physics and is solely determined by a closed contour in parameter space~\cite{Thouless,Niu,Sinitsyn07,Sinitsyn08,Sinitsyn07b}.  This geometric description can be used to optimize the pumping protocol~\cite{Flindt19} and may ensure a robust quantization of the pumped charge~\cite{Niu90,Hanggi10,Chernyak09}. 
Adiabatic pumps are important for a wide range of phenomena~\cite{Parrondo98,Ohkubo08,Sagawa11,Yuge12,Rahav08,Astumian07}, such as charge transport in nano-structures~\cite{Tabias}, heat transfer in molecular junctions~\cite{Dubi11}, and Brownian motors~\cite{Hanggi09,Hanggi05,Hanggi05a}.  Geometric pumping is also of interest in relation to stochastic thermodynamics, because it breaks the symmetry that leads to the steady-state fluctuation theorem~\cite{Seifert12,Esposito09,Saito08,Hayakawa17,Hayakawa19}.

For practical purposes, it would be useful to increase the driving frequency to produce a large output current. However, as the frequency is increased, non-adiabatic excitations tend to decrease the pumped charge, which in turn restricts the frequencies for which efficient charge pumping can be achieved. This situation resembles problems in quantum control theory, where fast driving speeds generally reduce both the fidelity and  robustness of a given operation~\cite{Nori05,Nori08}.
In this context, shortcuts to adiabaticity have recently been developed~\cite{STAreview,STAR} to realize adiabatic protocols in finite time. In particular, by using  counter-diabatic driving fields~\cite{STAreview,STAR,DR03,DR05,Berry09,Adolfo10,Jarzynski13,STAnonH}, a quantum system can be guided to follow a given adiabatic trajectory, and a desired operation can thereby be sped up. 

In this Letter, we develop a control scheme to speed up adiabatic pumping in classical stochastic systems. To this end, we construct a shortcut protocol which enables geometric pumping well beyond the adiabatic regime. We identify the target state of the system in the near-adiabatic regime and provide a systematic way of constructing the external control so that the system follows this target state even in the non-adiabatic regime. We emphasize that our method differs from existing shortcut protocols, such as the one for non-Hermitian systems~\cite{STAnonH}, since the required target state contains a non-adiabatic correction. As a specific application, we consider charge pumping through a quantum dot, Fig.~\ref{fig_rates}, for which we show that our method is robust against moderate fluctuations of the control field. We provide a geometric interpretation of the protocol and analyze the thermodynamic cost of implementing it. Our shortcut to geometric pumping can be realized with existing technology and, since it is universal (it does not rely on the adiabatic driving or on specific system properties), it may enable fast pumping of charge or heat in many different systems from physics, chemistry, and biology~\cite{Sinitsyn07,Sinitsyn07b,Sinitsyn08,Flindt19,Niu90,Hanggi10,Chernyak09,Parrondo98,Rahav08,Astumian07}.} 

{\it Charge pumping.---} We consider the classical stochastic dynamics of a system coupled to left and right reservoirs described by a master equation of the form 
$\del{}{t}\left|P(t)\right\rangle = L(t)\left| P(t)\right\rangle$. %\label{masterEQ}
Here, the vector $\left|P(t)\right\rangle =(p_{0}(t),p_{1}(t),\ldots)^{\mathrm{T}}$ contains the probabilities for the system to be in state $i=0,1,2,\dots$. The system is subjected to a periodic external drive, and the dynamics can be described by the rate matrix, $L(t)=L(t+\tau)$, where $\tau=2\pi/\Omega$ is the period of the drive. Below,
we discuss adiabatic pumping in a two-state system, because it is simple to analyse and relevant to experiments. However, the approach that we develop is more general, and it can be applied to systems with many states. The two-state system is important as it is equivalent to the orthodox model of a quantum dot in the strong Coulomb blockade regime, for which $p_{0}(t)$ and $p_{1}(t)$ would correspond to the probability of having $0$ or $1$ electrons on the dot. The rate matrix is now
\beq
L(t)=\left( \begin{array}{cc} -\Gamma^{+}_{\rm L}(t)-\Gamma^{+}_{\rm R}(t) & \Gamma^{-}_{\rm L}(t)+\Gamma^{-}_{\rm R}(t) \\ \Gamma^{+}_{\rm L}(t)+\Gamma^{+}_{\rm R}(t) & -\Gamma^{-}_{\rm L}(t)-\Gamma^{-}_{\rm R}(t) \end{array} \right), \label{rate}
\eeq
where $\Gamma^{\pm}_{\alpha}(t)$  are the time-dependent rates for an electron to tunnel on ($+$) or off ($-$) the quantum dot via the left or right reservoir, $\alpha={\rm L},{\rm R}$. In this case, the pumped charge per period into the right reservoir reads
\beq
\langle n\rangle =\int_0^\tau {\rm d}t \left[\Gamma^{-}_{\rm R}(t)p_{1}(t)-\Gamma^{+}_{\rm R}(t)p_{0}(t)\right], \label{avN}
\eeq
where the first term of the integrand is the average current running from the dot into the right reservoir, and the second term is the current in the opposite direction. To introduce a more general notation, we define the matrices
\beq
J_{+}(t)=\left( \begin{array}{cc} 0 & \Gamma^{-}_{\rm R}(t) \\ 0 & 0 \end{array} \right) \,\,\,\mathrm{and}\,\,\,
J_{-}(t)=\left( \begin{array}{cc} 0 & 0 \\ \Gamma^{+}_{\rm R}(t) & 0 \end{array} \right)\label{Jt},
\eeq
so that the pumped charge can be written as $\langle n\rangle=\int^{\tau}_{0}{\rm d}t \langle 1|J(t)|P(t)\rangle$, where $J(t)=J_{+}(t)-J_{-}(t)$ describes the current running into the right reservoir, and $\langle 1|=(1,1)$. This expression generalizes Eq.~(\ref{avN}) to systems with many states under an appropriate identification of $J(t)$.

\begin{figure*}
\begin{center}
\includegraphics[width=.85\textwidth]{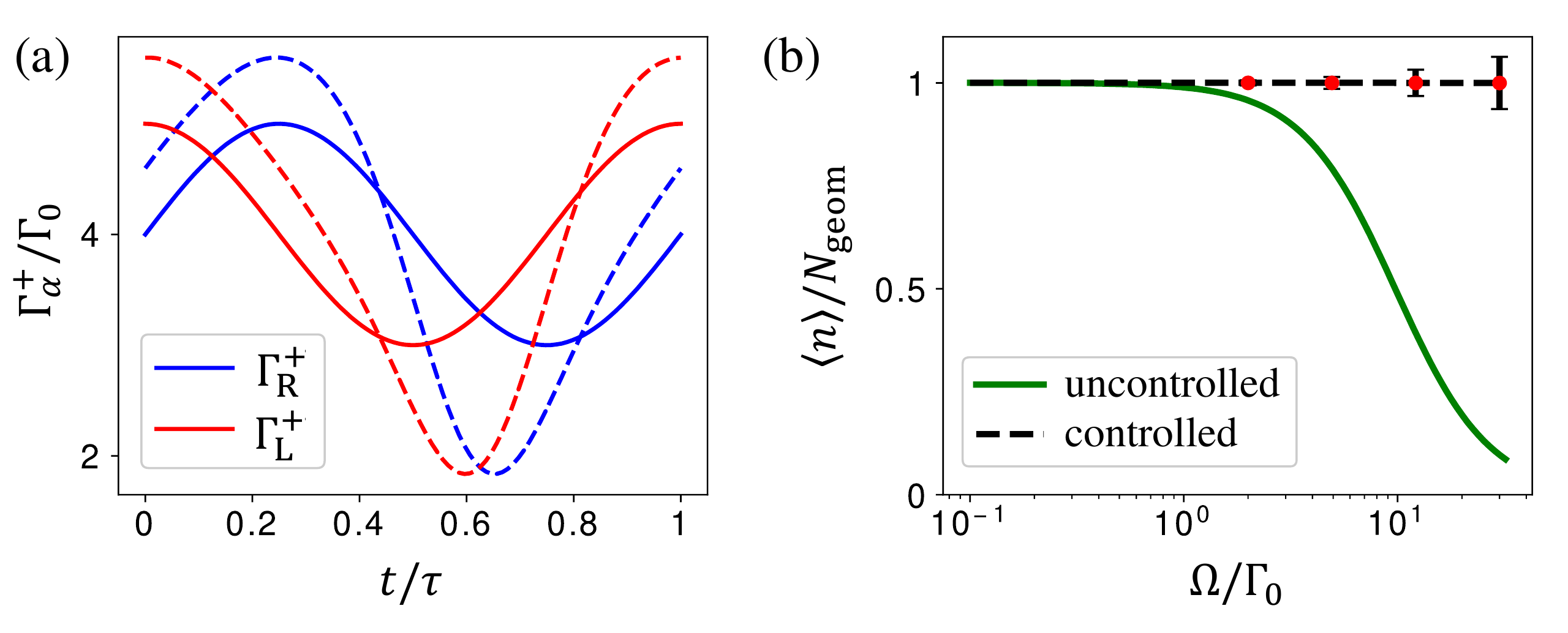}
\caption{{\revision{Transition rates and pumped charge. (a) The two time-dependent rates for the uncontrolled dynamics (full lines) and the corresponding modified rates (dashed lines). (b) Average pumped charge as a function of the driving frequency for the controlled and uncontrolled dynamics. The geometric description breaks down for the uncontrolled dynamics around $\Omega\simeq\Gamma_0$, where the pumped charge gets significantly reduced. In contrast, for the controlled dynamics, the pumped charge is given by the geometric contribution up until the breakdown frequency of around $\Omega_{\rm c}\simeq 32\Gamma_0$, where the modified transition rates become negative. The error bars are obtained by adding 3\% noise to the control field, showing that our protocol is robust against moderate fluctuations. The parameters used here are $A=4, R=1$, $\epsilon=0.5$, with $\Omega/\Gamma_{0}=25$ in (a) and $c=0.03$ in (b).} }  }
\label{fig_nav}
\end{center}
\end{figure*}

{\it Adiabatic pumping.---} To begin with, we consider charge pumping for slow drivings. We thus rewrite the master equation as $(L(t)-\del{}{t})\left| P(t)\right\rangle=0$ to evaluate the periodic state perturbatively in the frequency, $|P(t)\rangle = |\pi(t)\rangle + |\delta\pi(t)\rangle +\dots$, treating the time-derivative $-\del{}{t}$ as a perturbation of the instantaneous stationary state defined by $L(t)|\pi(t)\rangle=0$. Using standard perturbation theory, we then find
$|\delta\pi(t)\rangle = R(t)|\partial_{t}\pi(t)\rangle$, where $R(t)$ is the generalized inverse of $L(t)$~\cite{footnote1}, the normalization of $|P(t)\rangle$ implies that $\langle 1|\delta \pi(t)\rangle=0$, and we have defined $|\partial_t\pi(t)\rangle=\del{}{t}|\pi(t)\rangle$. Thus, we find for the near-adiabatic state the expression
\beq
|P_{\rm ad}(t)\rangle= |\pi(t)\rangle + R(t)|\partial_t\pi(t)\rangle .
\label{eq:target}
\eeq
Moreover, the pumped charge can be written as $\langle n \rangle_{\rm ad} =N_{\rm dyn}+N_{\rm geom}$, where $N_{\rm dyn}=\int^{\tau}_{0}{\rm d}t \langle 1|J(t)|\pi(t)\rangle$ is the period-averaged instantaneous current describing the dynamical steady-state contribution and
\beq
N_{\rm geom}=\int^{\tau}_{0}{\rm d}t\langle 1|J(t) R(t)|\partial_t\pi(t)\rangle=\oint_{\mathcal{C}} {\rm d}\bm{\Gamma} \!\cdot\! {\bm A} ({\bm \Gamma})  \label{Ngeom}
\eeq
is a purely geometrical contribution. To obtain Eq.~(\ref{Ngeom}), we use that the time-dependence of $J(t)$, $R(t)$, and $|\pi(t)\rangle$ enters implicitly through the transition rates $\Gamma^{\pm}_{\rm L,R}(t)$. Therefore, we can rewrite the integral over a period as an integral along the closed contour $\mathcal{C}$ in the parameter space $\bm{\Gamma}=(\Gamma^{+}_{\rm R},\Gamma^{+}_{\rm L},\Gamma^{-}_{\rm R},\Gamma^{-}_{\rm L})$.  We have also introduced the vector potential ${\bm A}({\bm \Gamma})=\langle 1|J[{\bm \Gamma}(t)]R[{\bm \Gamma}(t)] \frac{\partial}{\partial {\bm \Gamma}}|\pi[{\bm \Gamma}(t)]\rangle$, which is consistent with the classical analog of the Berry phase introduced by Sinitsyn and Nemenman in the context of full counting statistics~\cite{Sinitsyn07,Sinitsyn07b,Sinitsyn08}. Similar expressions have been obtained for pumped currents~\cite{Astumian07,Parrondo98,Rahav08} and entropy production~\cite{Sagawa11}, however, Eq.~(\ref{Ngeom}) is limited to adiabatic driving, and the geometric picture typically breaks down at higher frequencies.

{\it Shortcut to adiabatic pumping.---} We now develop a shortcut to geometric pumping beyond the adiabatic regime. To this end, we consider an external control that allows us to retain the target state (\ref{eq:target}) beyond the limit of slow driving. In this case, the non-Hermitian extension of the counterdiabatic technique cannot be used~\cite{STAnonH}, since the target state is not the instantaneous stationary state $|\pi(t)\rangle$. Instead, in the spirit of Ref.~\cite{Jarzynski13}, we note that the time evolution of the uncontrolled rate matrix $L(t)$ generates non-adiabatic excitations, such that the state cannot follow the target state (\ref{eq:target}). Specifically, for a short time-step $\delta t$, we have $|P_{\rm ad}(t)\rangle \rightarrow  \left(1+ L(t)\delta t\right) |P_{\rm ad} (t)\rangle  = |\pi(t+\delta t)\rangle + |\delta\pi(t)\rangle +O( \delta t^{2})$, since $|\partial_{t}\pi(t)\rangle=L(t)|\delta\pi(t)\rangle$, and this time-evolved state is different from the desired target state, $|P_{\rm ad}(t+\delta t)\rangle$. To circumvent this problem we construct an external control described by the matrix
\beq
L_{\rm cont}(t)=|\partial_{t}\delta\pi(t)\rangle\langle 1|,  \label{Lcont}
\eeq
which suppresses non-adiabatic excitations, so that the state of the system follows the desired target state as $|P_{\rm ad}(t)\rangle \rightarrow \left(1+\{ L(t)+L_{\rm cont}(t) \}\delta t \right) |P_{\rm ad}(t)\rangle = |P_{\rm ad}(t+\delta t)\rangle +O( \delta t^{2})$. Hence, with $|P_{\rm ad}(0)\rangle$ as the initial state, the solution to the master equation using the rate matrix $L(t)+L_{\rm cont}(t)$ is given by $|P(t)\rangle=|P_{\rm ad}(t)\rangle$ for all $t$. Importantly, the combined rate matrix $L(t)+L_{\rm cont}(t)$ must be physically meaningful at all times; specifically, all transition rates must remain non-negative. 

If the control is implemented without modifying the current operator~$J(t)$, we immediately find that the controlled dynamics reproduces the dynamical and the geometrical contribution to the pumped charge, i.~e.,
\beq
\langle n\rangle_{\rm cont}=\int^{\tau}_{0}{\rm d}t \langle 1 | J_{\rm cont}(t)|P_{\rm ad}(t)\rangle 
=N_{\rm dyn}+N_{\rm geom}, \label{controlledN}
\eeq
where $J_{\rm cont}(t)$ is the current operator with the control, and we have used $J_{\rm cont}(t)=J(t)$ to obtain the last equality in Eq.~(\ref{controlledN}). 
This condition is always satisfied, if we implement the control solely on the left side of the system. On the other hand, if the control is implemented on both sides, the current operator gets modified, $J_{\rm cont}(t)\neq J(t)$. However, Eq.~(\ref{controlledN}) still holds true, if we appropriately implement the control on both sides of the system by modifying the transition rates in $L(t)=L^{\rm L}(t)+L^{\rm R}(t)$ as $L^{\rm L}(t)\rightarrow L^{\rm L}(t)+(1-\epsilon) L_{\rm cont}(t)$ and $L^{\rm R}(t)\rightarrow L^{\rm R}(t)+\epsilon L_{\rm cont}(t)$. Here, $\epsilon$ is a free parameter that determines on which side the control is mainly implemented and $L^{\rm L(R)}(t)$ describes transitions into the left (right) reservoir. For $\epsilon=0$, the control is implemented solely on the left side of the system, and Eq.~(\ref{controlledN}) is satisfied. 
For $\epsilon=1$, a similar argument implies that the pumped charge per period from the left reservoir to the system satisfies $\langle n\rangle^{\rm L}_{\rm cont}=N_{\rm dyn}+N_{\rm geom}$. By using the conservation of the charge per period $\langle n\rangle^{\rm L}_{\rm cont}=\langle n\rangle_{\rm cont}$, Eq.~(\ref{controlledN}) is again valid. 
For $\epsilon$ between 0 and 1, we take a linear combination of the two limits to obtain Eq.~(\ref{controlledN}).

In summary, we find that our control enables geometric charge pumping beyond the limit of slow driving. The explicit form of the control (\ref{Lcont}) and its consequences for the charge pumping (\ref{controlledN}) are central results of this paper. We stress that the construction of the control (\ref{Lcont}) does not depend on the specific details of the system, including the number of states, and it is therefore universal in this sense. Moreover, while we here have focused on classical stochastic systems, it is clear that our control technique can also be applied to open quantum systems described by Markovian generalized master equations.

{\it Applications.---} \revision{To illustrate our control technique, we consider the two-state model in Eq.~(\ref{rate}) with the time-dependent rates, $\Gamma^{+}_{\rm L}(t)=\Gamma_{0}(A+R\cos\Omega t)$ and $\Gamma^{+}_{\rm R}(t)=\Gamma_{0}(A+R\sin\Omega t)$, plotted with full lines in Fig.~\ref{fig_nav}a, 
and $\Gamma^{-}_{\rm L}=\Gamma^{-}_{\rm R}=\Gamma_{0}$. The vector potential ${\bm A} ({\bm \Gamma})$ is shown in  Fig.~\ref{fig_rates}b together with the closed contour $\mathcal{C}$ defined by the rates. 
The dynamical contribution vanishes,  $N_{\rm dyn}=0$, while the geometrical term reads~\cite{Sinitsyn07}
\beq
  N_{\rm geom}=\frac{2\pi R^{2}}{\left[4(A+1)^{2}-2R^{2} \right]^{3/2}}. \label{sinexad}
\eeq
In Fig.~\ref{fig_nav}b we show the average pumped charge as a function of the driving frequency. At low frequencies, the driving is adiabatic, and the geometric description of the pumped charge is valid. However, as the frequency is increased and approaches the bare tunneling rate $\Gamma_0$, the driving becomes non-adiabatic, and the pumped charge becomes drastically reduced below the geometric value. 

To counteract the reduction of the pumped charge, we now implement our control.} The control matrix~(\ref{Lcont}) reads
\beq
L_{\rm cont}(t)=\gamma(t) \left(\begin{array}{cc} -1 & -1 \\ 1 & 1 \end{array}\right) ,
\eeq
where $\gamma(t)=\partial_{t}[\alpha(t)/\Gamma(t)]$ and $\alpha(t)=\partial_{t}[\Gamma^{-}(t)/\Gamma(t)]$, and we have defined $\Gamma^{\pm}(t)=\Gamma^{\pm}_{\rm L}(t)+\Gamma^{\pm}_{\rm R}(t)$ and $\Gamma(t)=\Gamma^{+}(t)+\Gamma^{-}(t)$. We have also used that $|\pi(t)\rangle =  ( \Gamma^{-}/\Gamma ,  \Gamma^{+}/\Gamma )^{\mathrm{T}}$ and $|\delta\pi(t)\rangle = (-\alpha/\Gamma, \alpha/\Gamma )^{\mathrm{T}}$. The control can be implemented by simply modifying the transition rates as $\Gamma^{\pm}_{\rm L}(t)\rightarrow \Gamma^{\pm}_{\rm L}(t)\pm (1-\epsilon)\gamma(t)$ and $\Gamma^{\pm}_{\rm R}(t)\rightarrow \Gamma^{\pm}_{\rm R}(t)\pm \epsilon\gamma(t)$, where $\epsilon$ can be between 0 and 1. 

\revision{Figure~\ref{fig_nav}a shows two of the modified rates (dashed lines) for a fixed driving frequency, while the resulting pumped charge is plotted as a function of the frequency in panel~b. Using the external control, we find that the geometric value is maintained for frequencies that are more than an order of magnitude larger than the breakdown frequency for the uncontrolled dynamics. At very large frequencies, the modified transition rates become negative and the control can no longer be implemented. The breakdown frequency $\Omega_{\rm c}$ can be estimated as $\Omega_{\rm c}= \Gamma_{0} \left\{2[1+\sqrt{2}(1+A)/R]\right\}^{3/2}\simeq 32\Gamma_{0}$ for $A=4$ and $R=1$, which agrees well with our numerical results.

The robustness of our protocol is important for practical realizations. To gauge the stability of our scheme, we show in Fig.~\ref{fig_nav}b the standard deviation of the pumped charge, obtained by adding random Gaussian noise $\xi_{t}$ to the control field as $\gamma(t)\rightarrow (1+c\xi_{t})\gamma(t)$, where $c$ is the noise strength and $\av{\xi_{t}}=0$ and $\av{\xi_{t}\xi_{t'}}=\delta(t-t')$. The smallness of the error bars demonstrates that our scheme is stable against moderate fluctuations.

{\it Geometric interpretation.---}  The control itself has an interesting geometric interpretation. To see this, we define the distance between two states $|p\rangle$ and $|q \rangle$ with components $p_{i}$ and $q_{i}$ as $\mathcal{L}(p,q)=\sqrt{\sum_{i}| p_{i}-q_{i}|^{2}}$. Now, by considering the infinitesimal distance between two neighboring states $\mathcal{L}(p(t+\delta t),p(t))=\sqrt{g_{tt}[p(t)]}\delta t+O(\delta t^{2})$, we are led to define the metric as  $g_{tt}[p(t)]=\sum_{i}|\partial_{t}p_{i}(t)|^{2}$~\cite{Caves94}. We then see that $L_{\rm cont}$ is related to the geometry of the correction to the stationary state as
\beq
||L_{\rm cont}||_{\rm F}=\sqrt{N\sum_{i} | \partial_{t}\delta\pi_{i}(t) |^{2}}=\sqrt{Ng_{tt}[\delta\pi(t)]}, \label{Lnorm}
\eeq
where $N$ is the number of states of the system and  $||A||_{\rm F}=\sqrt{\sum_{i,j}|a_{ij}|^{2}}$ is the Frobenius matrix norm. This relation indicates that, if $|\delta\pi(t)\rangle$ depends strongly on time, a large intensity of the control is required to suppress the nonadiabatic excitations.

Equation~(\ref{Lnorm}) should be contrasted with the non-Hermitian control field of Ref.~\cite{STAnonH}, which is given as $L_{\rm cd}(t)=|\partial_{t}\pi(t)\rangle\langle 1|$, and which is related to the geometry of the stationary state as $||L_{\rm cd}||_{\rm F}=\sqrt{Ng_{tt}[\pi(t)]}$. In this respect, our control matrix can be regarded as a next-order non-adiabatic generalization of $L_{\rm cd}$. For the two-state case, we have $\sqrt{g_{tt}[\pi(t)]}=\sqrt{2}|\alpha(t)|$ and $\sqrt{g_{tt}[\delta\pi(t)]}=\sqrt{2}|\gamma(t)|$. A similar connection has been discussed between the counterdiabatic Hamiltonian and the Fubini-Study metric in quantum systems, which has led to several speed-limits and trade-off relations~\cite{Polkovnikov19,Santos15,Campbell17,Funo17}. Thus, we  expect that similar universal relations may also exist for the geometric pumping considered here.
}

{\it Thermodynamic cost.---} Finally, we compare the entropy production for the controlled dynamics with its adiabatic counterpart. Since our system is interacting with two reservoirs, a steady-state heat current (housekeeping heat) can be generated even in the stationary state. We therefore consider the Hatano-Sasa entropy production, which is the entropy production of the reservoirs after subtracting the entropy that is generated by the stationary dissipation. The Hatano-Sasa entropy production for a stationary cycle is defined as~\cite{HS}
\beq
\Sigma_{\rm HS}=\int^{\tau}_{0}{\rm d}t \sum_{ji}W_{ji}p_{i}\ln\left( \frac{p^{\rm SS}_{j}}{p^{\rm SS}_{i}} \right) \geq 0,
\eeq
where $W_{ji}$ is the $j,i$'th component of the rate matrix~$W$ and $p^{\rm SS}_{i}$ is the $i$'th component of the corresponding instantaneous stationary state.

For the uncontrolled dynamics, we have $p_{i}=\pi_{i}+\delta\pi_{i}$ as expressed by Eq.~(\ref{eq:target}), and with $W=L$ and $p^{\rm SS}_{i}=\pi_{i}$, we find that the entropy production vanishes, $\Sigma_{\rm ad}=0$~\cite{footnote2}. By contrast, the control required to mimic the near-adiabatic dynamics generates a finite amount of entropy. For the two-state case, we have $W_{01}=\Gamma^{-}-\gamma$, $W_{10}=\Gamma^{+}+\gamma$, $p_{0}=(\Gamma^{-}-\alpha)/\Gamma$, and %$p_{1}=(\Gamma^{+}+\alpha)/\Gamma$, 
$p^{\rm SS}_{0}=(\Gamma^{-}-\gamma)/\Gamma$. The Hatano-Sasa entropy production then becomes
\beq
\Sigma_{\rm cont}
=\int^{\tau}_{0}{\rm d}t (\alpha-\gamma)\ln \left( \frac{1-\gamma/\Gamma^{-}}{1+\gamma/\Gamma^{+}}\right)-\int^{\tau}_{0}{\rm d}t \gamma \ln \left( \frac{\Gamma^{-}}{\Gamma^{+}} \right).
\label{controlEntpro}
\eeq
The strength of the control, $\gamma\propto \Omega^{2}$, vanishes at low frequencies and we recover $\Sigma_{\rm cont}\rightarrow \Sigma_{\rm ad}=0$. On the other hand, at large frequencies, the control can no longer be implemented and the entropy production diverges. Hence, we interpret the  entropy production in Eq.~(\ref{controlEntpro}) as the thermodynamic cost of implementing our control.

{\it Conclusions.---} \revision{We have developed a shortcut to geometric pumping in classical stochastic systems. Going beyond existing protocols for non-Hermitian systems, our shortcut makes it possible to recover the near-adiabatic dynamics much beyond the adiabatic regime and thereby maintain the geometrical description of the pumped charge. The control protocol can be implemented by modifying the transition rates of the uncontrolled system, and it is robust against moderate fluctuations of the control field. Our work opens several promising avenues for further developments. Experimentally, our control can be realized in systems from physics, chemistry, and biology. Theoretically, it would be interesting to explore possible speed-limits and trade-off relations, similarly to those that have been formulated for counterdiabatic Hamiltonians in quantum systems. Extending our ideas to adiabatic pumping in quantum systems constitutes another interesting line of research \cite{Brouwer98}.}

\begin{acknowledgements}
{\it Acknowledgements.---} We thank K.~Brandner, E.~Potanina, K.~Saito, for useful discussions, and K.~Fujii, H.~Hayakawa, Y.~Hino, and K.~Takahashi for drawing our attention to their recent pre-print \cite{fujii2019nonadiabatic}. K.~F.~was supported by the JSPS KAKENHI Grant Number JP18J00454.  N.~L.~acknowledges partial support from JST PRESTO through Grant No.~JPMJPR18GC. F.~N. is supported in part by the: MURI Center for Dynamic Magneto-Optics via the Air Force Office of Scientific Research (AFOSR) (FA9550-14-1-0040),
Army Research Office (ARO) (Grant No.~W911NF-18-1-0358), Asian Office of Aerospace Research and Development (AOARD) (Grant No.~FA2386-18-1-4045),
Japan Science and Technology Agency (JST) (via the Q-LEAP program, and the CREST Grant No.~JPMJCR1676), NTT Physics and Informatics (PHI) Lab, Japan Society for the Promotion of Science (JSPS) (JSPS-RFBR Grant No.~17-52-50023, and JSPS-FWO Grant No.~VS.059.18N). F.~N.~and N.~L.~also acknowledge support from the RIKEN-AIST Challenge Research Fund, and by grant number FQXi-IAF19-06 from the Foundational Questions Institute Fund, a donor advised fund of Silicon Valley Community Foundation. C.~F.~was supported by the Academy of Finland (projects No.~308515 and 312299).
\end{acknowledgements}


\begin{thebibliography}{99}
\bibitem{footnote1} A generalized inverse is not uniquely defined, but we here fix our choice by imposing the condition $\langle 1|R(t)=0$, which implies $\langle 1|\delta \pi(t)\rangle=0$. In that case, the generalized inverse reads $R(t)=\sum_{\mu}\phi_{\mu}^{-1}(t)|\phi_{\mu}(t)\rangle\langle\widetilde{\phi}_{\mu}(t)|$, where $\phi_{\mu}(t)$, $\langle \widetilde{\phi}_{\mu}(t)|$, and $|\phi_{\mu}(t)\rangle$ are respectively the nonzero eigenvalues, left eigenvectors, and right eigenvectors of $L(t)$, i.~e., $L(t)=0\cdot |\pi(t)\rangle\langle 1|+\sum_{\mu}\phi_{\mu}(t)|\phi_{\mu}(t)\rangle\langle\widetilde{\phi}_{\mu}(t)|$. For the two-state system, we have
\beq
R(t)=\frac{1}{\Gamma^{2}(t)}\left( \begin{array}{cc} -\Gamma^{+}(t) & \Gamma^{-}(t) \\ \Gamma^{+}(t) & -\Gamma^{-}(t) \end{array} \right),
\eeq
where $\Gamma^{\pm}(t)=\Gamma^{\pm}_{\rm L}(t)+\Gamma^{\pm}_{\rm R}(t)$ and $\Gamma(t)=\Gamma^{+}(t)+\Gamma^{-}(t)$.
\bibitem{footnote2} We first use the normalization condition $\sum_{j}L_{ji}=0$ and the property of the stationary state $\sum_{i}L_{ji}\pi_{i}=0$ and obtain $\Sigma_{\rm ad}=\int^{\tau}_{0}dt \sum_{ji}L_{ji}\delta\pi_{i}\ln \pi_{j}$. By noting that the integrand becomes a total derivative, i.e., $\sum_{ji}L_{ji}\delta\pi_{i}\ln \pi_{j}=\sum_{j}(\partial_{t}\pi_{j})\ln\pi_{j}=\partial_{t}(\sum_{j}\pi_{j}\ln\pi_{j})$, we have $\Sigma_{\rm ad}=0$. %Here, we use $|\partial_{t}\pi\rangle=L|\delta\pi\rangle$ and the normalization condition $\sum_{j}\partial_{t}\pi_{j}=0$.
\end{thebibliography}
\end{document}